\documentclass[prl,twocolumn,showpacs,preprintnumbers,amsmath,amssymb,superscriptaddress]{revtex4}

\usepackage{graphicx}
\usepackage{dcolumn}
\usepackage{bm}
\usepackage{amsmath}	
\begin{document}
\draft
\newcommand{\lw}[1]{\smash{\lower2.ex\hbox{#1}}}

\title{Magnetism Localization in Spin-Polarized One-Dimensional
Anderson--Hubbard Model}

\author{M.~Okumura}
\email{okumura.masahiko@jaea.go.jp}
\affiliation{CCSE, Japan Atomic Energy Agency, 6--9--3 Higashi-Ueno,
Taito-ku, Tokyo 110--0015, Japan}
\affiliation{CREST (JST), 4--1--8 Honcho, Kawaguchi, Saitama 332--0012,
Japan}
\author{S.~Yamada} 
\email{yamada.susumu@jaea.go.jp}
\affiliation{CCSE, Japan Atomic Energy Agency, 6--9--3 Higashi-Ueno,
Taito-ku, Tokyo 110--0015, Japan}
\affiliation{CREST (JST), 4--1--8 Honcho, Kawaguchi, Saitama 332--0012,
Japan}
\author{N.~Taniguchi}
\email{taniguch@sakura.cc.tsukuba.ac.jp}
\affiliation{Institute of Physics, University of Tsukuba, Tennodai,
Tsukuba 305--8571, Japan}
\author{M.~Machida}
\email{machida.masahiko@jaea.go.jp}
\affiliation{CCSE, Japan Atomic Energy Agency, 6--9--3 Higashi-Ueno,
Taito-ku, Tokyo 110--0015, Japan}
\affiliation{CREST (JST), 4--1--8 Honcho, Kawaguchi, Saitama 332--0012,
Japan}

\date{\today}

\begin{abstract} 
 In order to study an interplay of disorder, correlation, and spin 
 imbalance on antiferromagnetism, we systematically explore the ground
 state of one-dimensional spin-imbalanced Anderson--Hubbard model by
 using the density-matrix renormalization group method. We find that
 disorders localize the antiferromagnetic spin density wave induced by
 imbalanced fermions and the increase of the disorder magnitude shrinks
 the areas of the localized antiferromagnetized regions. Moreover, the
 antiferromagnetism finally disappears above a large disorder. These
 behaviors are observable in atomic Fermi gases loaded on optical
 lattices and disordered strongly-correlated chains under magnetic
 field.
\end{abstract}
\pacs{71.10.Fd, 71.10.Pm, 71.23.-k, 03.75.Ss}

\maketitle

An atomic Fermi gas loaded on an optical lattice (FGOL) has been one of
the most active target in the atomic gas field \cite{OLFermi} since the
successful observation of the superfluid-insulator transition in the
Bose counterpart \cite{OLBose}. In FGOL, its interaction tunability 
associated with the Feshbach resonance and lattice formation flexibility
due to optical operations offer a great opportunity to systematically
study the Hubbard model \cite{stripe} and its extended versions
\cite{Gao,Okumura}. Thus, FGOL is expected to be a promising
experimental reality to resolve a wide range of controversial issues in
condensed matter physics \cite{OLtheory}. 

The pairing mechanism in High-$T_{\rm c}$ superconductors has been one
of the mostly debated issues for the latest two decades. The
superconductivity emerges by doping holes via chemical substitutions on
the Mott-insulator mother phase \cite{dopedMott}. This drastically
complicates theoretical studies because the pairing occurs on the
strongly-correlated stage influenced by disorders. This fact stimulates
systematic studies for the Anderson--Hubbard (AH) model
\cite{Anderson,Hubbard,1DAH,1DAHring}. FGOL is an good experimental
testbed to solve the complicated problem \cite{Gao,Okumura}.  

In addition to such a practical motivation from the condensed matter
side, FGOL inspires its own unique interests which lead to new
frontiers. Its typical example is an arbitrary spin imbalance, which has 
been the most intensive target for a recent few years in atomic Fermi
gases \cite{ImbalanceEx}. Therefore, in this paper, we investigate the
spin imbalance effects on the AH model. This will be one of the most
fruitful target as well as the balanced case for the future FGOL
experiments. Moreover, we point out that its weak imbalance case has a
real counterpart, which corresponds to disordered strongly-correlated
materials under magnetic field \cite{MottHeisenberg}. Thus, we examine
the model from the weak imbalance side. 

The spin imbalance is believed to bring about a spatially inhomogeneous  
pairing like Fulde--Ferrel--Larkin--Ovchinikov state \cite{FFLO} in
attractive Hubbard model \cite{NumFF}. This is a widely spread idea,
while the imbalance effects have been very little investigated in the 
repulsive case. We therefore concentrate on antiferromagnetic phases
characteristic to the repulsive Hubbard model and study how not only the
imbalance but also the disorder magnitude affect the phases. In this
paper, we focus on one-dimensional (1D) repulsive polarized
AH model at the half-filling by using the density matrix renormalization
group (DMRG) method \cite{White,DMRGreview}. The highlight in this paper
is that disorders localize antiferromagnetic phases accompanied by a
phase separation from the non-magnetized one and finally eliminate the
antiferromagnetism in a strong disorder range. These results are
directly observable in FGOL's by using the atomic-density profile probe
which is presently the most standard technique (for the experimental
setup, see Ref.~\cite{Okumura}).

The Hamiltonian of the 1D AH model is given by
\begin{equation}
 H_{\rm AH} = -t \sum_{\langle i,j \rangle,\sigma} c^\dag_{\sigma i}
 c^{\phantom{\dag}}_{\sigma j} + \sum_{i, \sigma} \epsilon_i n_{\sigma
 i} + \sum_{i} U n_{\uparrow i} n_{\downarrow i} \, ,
 \label{Hamiltonian} 
\end{equation}
where $\langle i ,j \rangle$ refers to the nearest neighbors $i$ and
$j=i\pm1$, $t$ is the hopping parameter between the nearest neighbor
lattice sites, $U$ is the on-site repulsion, $c^{\phantom{\dag}}_{\sigma
i}$ ($c_{\sigma i}^\dag$) is the annihilation- (creation-) operator and 
$n_{\sigma i} (\equiv c^\dag_{\sigma i} c^{\phantom{\dag}}_{\sigma i})$
is the site density operator with spin index
$\sigma=\uparrow,\downarrow$, and the random on-site potential
$\epsilon_i$ is chosen by a box probability distribution ${\cal P}
(\epsilon_i) = \theta ( W/2 - |\epsilon_i|)/W$, where $\theta (x)$ is
the step function and the parameter $W$ the disorder strength
magnitude. In DMRG calculations, the number of states kept ($m$) is
$500$--$700$ and these numbers ($m \ge 500$) are enough for the most
cases. In the calculations, we apply the open boundary condition except
for the comparison with the periodical condition and measure the site
matter and spin density of fermions as $n_{\uparrow i} \pm n_{\downarrow
i}$. 

\begin{figure}
\includegraphics[scale=0.36]{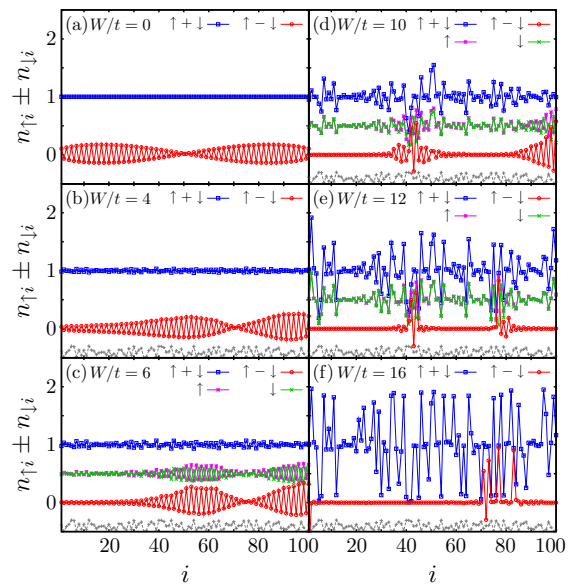}
\caption{\label{fig1} The randomness magnitude $W$ dependence of the
 matter and spin density profiles ($n_{\uparrow i} \pm n_{\downarrow i}$
 respectively) at the half-filling for $U/t=10$ with $(N_\uparrow,
 N_\downarrow)=(51,49)$. A profile of the selected random potential is
 depicted on the bottom of each figure in an arbitrary unit (grey dashed
 line). In figures (c)--(e), up and down spin density profiles are
 shown. For all calculations, $m$ is 500.}
\end{figure}

Let us show DMRG calculation results on the spin imbalanced AH model
(\ref{Hamiltonian}). Figure \ref{fig1} displays a randomness amplitude
($W/t$) dependence of the on-site matter and spin density profile for
the number of the total sites $L=100$, the number of the spin-up and
spin-down fermions $(N_\uparrow, N_\downarrow)= (51,49)$ (half-filling),
and $U/t=10$. In this case, two up-spin fermions do not have their
(down-spin) partners. The matter density profile is almost flat for $W
\le U$ as seen in Figs.~\ref{fig1}(a)--\ref{fig1}(c), while it is
drastically disturbed for $ W \ge U $. This is completely the same as
that of the well-known balanced case. However, the spin density profile
is significantly different. Firstly, in the clean case as shown in
Fig.~\ref{fig1}(a) ($W/t =0$), one finds an antiferromagnetic spin
density wave (ASDW) whose periodicity is found to be inversely
proportional to $N_\uparrow - N_\downarrow$ [e.g., see 
Figs.~\ref{fig3}(h) and \ref{fig3}(i), in which $N_\uparrow -
N_\downarrow = 2$ and $4$, respectively]. Here, it is noted that any
ASDW phases are never observed in the perfectly balanced cases
irrespective of the presence of randomness \cite{note1}. This clearly
indicates that the imbalance is responsible for the ASDW
phase. Secondly, as one increases the disorder strength, a part of the
ASDW (amplitude) ``locally'' vanishes and the depressed regions expand
as seen in Figs.~\ref{fig1}(b) and \ref{fig1}(c). This tendency becomes
remarkable when $W$ exceeds over $U$ as seen in Figs.~\ref{fig1}(d) and
\ref{fig1}(e), in which two ASDW phases are localized and isolated each
other. This isolation can be explained by the complete localization of
two excess up-spin fermions for $W > U$ since the ASDW can be created
only at the localized spots of the excess up-spin particles. In fact,
the change in the up and down spin profiles in
Fig.~\ref{fig1}(c)--\ref{fig1}(e) supports the idea. The further
increase of $W$ diminishes the antiferromagnetic structure as seen in
Fig.~\ref{fig1}(f), in which the localized structure is characterized by
positive peaks instead of the staggered (plus-minus) moment
alternation. This is because the strong randomness fully dominates over
the other effects. However, we note here that it is difficult in this
large disorder range to judge whether the result [Fig.~\ref{fig1}(f)] is
the true ground-state or not. The reason is that in this strong $W$
range tiny changes of $W$ (e.g., $W+\delta W$) give entirely different
spin-density distributions in non-continuous manner, which is not
observed when $0 < W/t \le 14$. Generally, it is well-known in strongly
glassy situations that a tiny change in calculation parameters results
in a drastic different consequence. Thus, we expect that there are a lot
of local minima in this strong disorder range. Fig.~\ref{fig1}(f) is a
localization profile selected among those minima.
 
\begin{figure}
\includegraphics[scale=0.36]{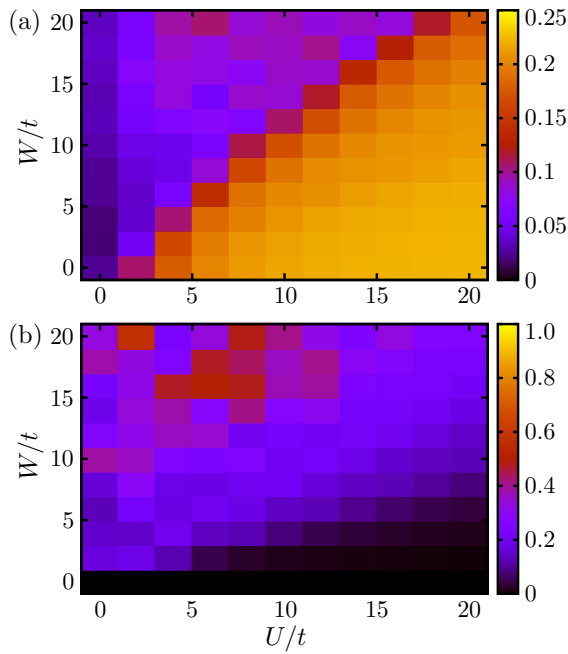}
\caption{\label{fig2} Contour plots of the value of (a) $S(U,W)$
 [Eq.~(\ref{SUW})] and (b) $D(U,W)$ [Eq.~(\ref{DUW})] with $L=100$ and 
 $(N_\uparrow,N_\downarrow)=(101,99)$ at the half-filling in a range of
 $U$ $(0 \le U \le 20)$ and $W$ $(0 \le W \le 20)$. The step value for
 both $U/t$ and  $W/t$ is 2. For all calculations, $m$ is 500.} 
\end{figure}

\begin{figure}
\includegraphics[scale=0.36]{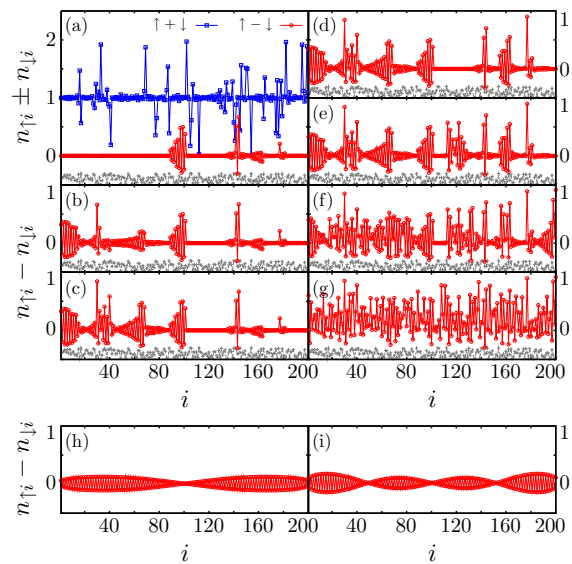}
\caption{\label{fig3} The polarization dependence of the spin density
 profiles [(a)--(g)] with $U/t=W/t=20$ and that without random potential
 [(h) and (i)] at $U/t=20$. The last two figures (h) and (i) without the
 randomness are displayed for a comparison with disordered cases (a) and
 (b) respectively. The number of up and down spin particles
 $(N_\uparrow,N_\downarrow)$ are (a) $(101,99)$, (b) $(102,98)$, (c)
 $(103,97)$, (d) $(104,96)$, (e) $(105,95)$, (f) $(110,90)$, (g)
 $(120,80)$, (h) $(101,99)$, and (i) $(102,98)$, respectively. The
 random potential shape are displayed on the bottom of the figures
 (a)--(g) (gray dashed lines). For all calculations, $m$ is 600.} 
\end{figure}

In order to characterize the spin density profile in a wide range of $U$
and $W$, we define the following function, 
\begin{equation}
 S(U,W)= \left\langle \sum_{i=1}^{L} \frac{\left| n_{\uparrow i}
 (\epsilon,U,W) - n_{\downarrow i} (\epsilon,U,W) \right|}{L}
 \right\rangle_\epsilon , \label{SUW} 
\end{equation}
where $n_{\sigma i}(\epsilon,U,W)$ is the local site density under a
random potential symbolically specified by $\epsilon$ at a certain set
of $U$ and $W$, and $\langle \cdot \rangle_\epsilon$ means an algebraic
average for various random realizations. From the expression, it is
found that $S(U,W)$ gives an indicator how large the spin moment
develops on each site. If the staggered moment widely grows, then
$S(U,W)$ gives a relatively large value. Thus, a map of $S(U,W)$ in a
wide range of $U$ and $W$ is expected to clarify an interplay of $U$ and
$W$ on the ASDW phase localization. Figure \ref{fig2}(a) shows a contour
plot of $S(U,W)$, which is averaged over ten realizations of random
potentials for $L=100$ and $(N_\uparrow,N_\downarrow)=(51,49)$. In this
figure, when one increases $W$ along a fixed $U/t$ line (e.g., $U/t=10$
line) from $W/t=0$ to $20$, it is found that the averaged moment of ASDW
very slowly decreases inside the region $W < U$. This is consistent with
the spin density profile as seen in Figs.~\ref{fig1}(a)--\ref{fig1}(c) 
in which the areas of ASDW phases slowly diminish with increasing
$W$. When $W$ exceeds over $U$, the variation of $S(U,W)$ suddenly
changes to a fast suppression. This reflects the change in the
localization of the excess up-spin particles as seen in
Figs.~\ref{fig1}(c) and \ref{fig1}(d). 

In order to qualify the present randomness averaging on
Fig.~\ref{fig2}(a), we introduce the following function
\begin{equation}
 D(U,W) = \sqrt{\left\langle \left\{ [S_\epsilon (U,W) / S (U,W)] - 1 
 \right\}^2 \right\rangle_\epsilon} \, . \label{DUW}
\end{equation}
This corresponds to a standard deviation on the randomness average. 
Fig.~\ref{fig2}(b) is a contour map of $D(U,W)$ in the same range as
Fig.~\ref{fig2}(a). One finds that in a small $W/t$ range $D(U,W)$ shows
a very small value (almost zero) and $D(U,W)$ reaches about $0.2$ around
$W=U+3t \sim 4t$ line when increasing $W/t$ at a constant $U/t$. Thus,
the averaged values in Fig.~\ref{fig2}(a) are sufficiently qualified
except for above $W=U+3t$ line.  This result means that the qualitative
change as observed around $W=U$ in Fig.~\ref{fig2}(a) is not a side
effect associated with the averaging but an essential feature in this
system. 

Next, we investigate the polarization strength dependence of the spin
density profile. Figures \ref{fig3}(a)--\ref{fig3}(g) display the
correspondent results made at the half-filling in $L=200$, $U/t=18$ and 
$W/t=20$. In the range of these parameters, clear localization of the
ASDW phases can be observed with the phase separation from the
non-magnetized phases in a slight polarized case [e.g., see
Fig.~\ref{fig1}(d)]. Firstly, Fig.~\ref{fig3}(a) displays the charge and 
spin density distributions in $(N_\uparrow,N_\downarrow)=(101,99)$. One
finds two magnetized regions, in ASDW phases are localized with the
localization of two extra fermions. The slight increase of the
polarization [$(N_\uparrow,N_\downarrow)=(102,98)$] increases the number 
of the magnetized regions as seen in Fig.~\ref{fig3}(b). Here, we note
that the magnetized regions formed in the less polarized case as 
Fig.~\ref{fig3}(a) are kept. This is in contrast to the clean systems
($W/t=0$) as shown in Figs.~\ref{fig3}(h) and \ref{fig3}(i) whose
polarizations are the same as Figs.~\ref{fig3}(a) and \ref{fig3}(b),
respectively. This is a typical feature characteristic to disordered
systems, in which memory effects can be frequently observed. The further
increase of the polarization extends the magnetized regions as shown in
Figs.~\ref{fig3}(c)--\ref{fig3}(g). In Fig.~\ref{fig3}(g), the
magnetized (ASDW) regions cover all sites, which are almost positively
polarized although the alternation profile still remains. 

\begin{figure}
\includegraphics[scale=0.36]{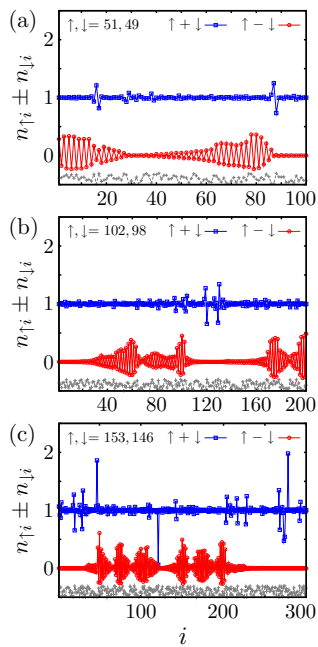}
\caption{\label{fig4} The size dependence of the matter and spin density
 profiles. The same polarization ratio is kept. (a) $L=100$ case with 
 $(N_\uparrow,N_\downarrow)=(51,49)$, (b) $L=200$ case with
 $(N_\uparrow,N_\downarrow)=(102,98)$, and (c) $L=300$ case with
 $(N_\uparrow,N_\downarrow)=(153,147)$, respectively. $U/t=W/t=20$ in
 all cases. The random potential profiles are shown on the bottom of
 each figure (grey dashed lines). For all calculations, $m$ is 600.} 
\end{figure}

We investigate a system size dependence of this magnetism localization
to confirm that it is not a small size effect. In Figures
\ref{fig4}(a)--\ref{fig4}(c), we show the charge and spin density
distributions at $U/t=W/t=20$ in $L=100$, $200$, and $300$ with
$(N_\uparrow,N_\downarrow)=(51,49)$, $(102,98)$, and $(153,147)$, 
respectively. We clearly find that these all cases exhibit the same
qualitative behavior, i.e., the magnetized regions are localized with
the separation from the non-magnetized regions. From these figures, we
find that the observed magnetism localization is an intrinsic effect.  

\begin{figure}
\includegraphics[scale=0.36]{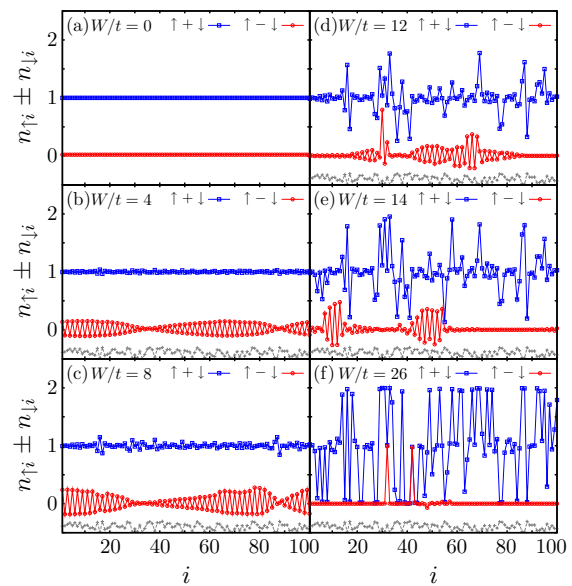}
\caption{\label{fig5} The randomness $W$ dependent matter and spin
 density profiles ($n_{\uparrow i} \pm n_{\downarrow i}$ respectively)
 with $L=100$, $(N_\uparrow,N_\downarrow)=(51,49)$, and $U/t=10$, under
 a random potential depicted on the bottom of each figure in arbitrary
 unit (grey dashed line). The periodic boundary condition is
 employed. For all calculations, $m$ is 700.}  
\end{figure}

Finally, we re-examine the model (\ref{Hamiltonian}) under the periodic
boundary condition to check the effect of the boundary
condition. Figures \ref{fig5}(a)--\ref{fig5}(f) show the $W$ dependent
charge and spin density profiles in the periodic condition. In
Fig.~\ref{fig5}(a) ($W/t=0$), we find complete flat distributions in
both the charge and spin densities. They are characteristic to the
periodic boundary condition in which the excess fermions fully
distribute homogeneously. This is in contrast to the open boundary
condition [compare it with Fig.~\ref{fig1}(a)]. When the randomness is
added into the system, the ASDW phases are induced as seen in
Figs.~\ref{fig5}(b) and \ref{fig5}(c). Thus, one finds that the ASDW
phase requires two conditions, i.e., the imbalance and the translational
symmetry breaking. When the randomness strength increases, one finds
that the amplitude of the ASDW increases [see Figs.~\ref{fig5}(b) and 
\ref{fig5}(c)]. This implies that the localization of extra two fermions  
proceeds with increasing $W/t$. At $W/t=12$ and $14$ [see
Figs.~\ref{fig5}(d) and \ref{fig5}(e) for $W>U$], we find the phase
separation from the non-magnetized phases, which is the same as the open
boundary case. Further increase of $W/t$ brings about more tight
localization [Fig.~\ref{fig5}(e)] and disappearance of the staggered
moment profile [Fig.~\ref{fig5}(f)]. This is also the same as the open
boundary case. 

In conclusion, we systematically studied the polarized AH model at the
half-filling and found that the disorder localizes the ASDW phases
induced by the excess fermions. As the randomness strength increases,
the areas of the localized ASDW phases shrink with the expansion of the
non-magnetized areas, and the antiferromagnetism finally vanishes. These
novel disorder effects on polarized strongly-correlated systems are
observable in not only 1D FGOL's \cite{Okumura} but also
strongly-correlated disordered chains under the magnetic field. 

The authors wish to thank H.~Aoki, T.~Deguchi, K.~Iida, T.~Koyama,
H.~Matusmoto, Y.~Ohashi, T.~Oka, S.~Tsuchiya, and Y.~Yanase for
illuminating discussion. 
The work was partially supported by Grant-in-Aid for Scientific Research
(Grant No.~18500033) and one on Priority Area ``Physics of new quantum
phases in superclean materials'' (Grant No.~18043022) from MEXT, Japan.



\begin{references}
\bibitem{note1}
In this system, $N_\uparrow$ and $N_\downarrow$ are noted to be
conserved respectively. They are also essentially conserved in FGOL.
\end{references}
\end{document}